\documentclass[11pt]{article}
\usepackage{epsfig}

\author{S. Sivasubramanian, Y. N. Srivastava$^\dagger $, A. Widom \\
{\it Physics Department, Northeastern University, Boston MA USA} \\
{\it and}\\
{\it $^\dagger $Physics Department \& INFN, University of Perugia, Perugia IT}}

\title{Landau Ghosts and Anti-Ghosts in Condensed Matter and High Density Hadronic Matter}

\begin{document}
\maketitle
\vskip .5cm
\centerline{Abstract}
\bigskip
\par \noindent
{\it It is observed that the ``ghost'' (originally discovered by Landau
in quantum electro-dynamics) and its counterparts in other theories
are indeed ubiquitous as they occur in a one-loop approximation to any
conventional (unbroken) gauge theory. The mechanism is first
exposed in its generality via the Dyson equation and a simple but
explicit example in condensed matter is provided through the
static Clausius-Mossotti and its dynamic counterpart the
Lorenz-Lorentz equation. The physical phase transition phenomenon
associated with it is found to be super-radiance. We verify
quantitatively that water (and many other polar liquids) are
indeed super-radiant at room temperature. In quantum
chromo-dynamics on the other hand, we encounter, thanks to
asymptotic freedom, an ``anti-ghost'' which is closely associated
with color confinement. Thus, in QCD, free quarks and glue exist
in a super-radiant phase and hadronic matter in the normal phase.}

\vfill \eject

\section{Introduction}
  It was discovered many decades ago\cite{Landau}, that the dimensionless
coupling constant in QED, \begin{math}\alpha_0 = e_0^2/\hbar
c\end{math}, after 1-loop renormalization becomes momentum transfer
\begin{math}Q^2 \end{math} dependent
\begin{math}\alpha(Q^2)\end{math} and that it
has a pole at a (space-like \begin{math} Q^2>0 \end{math})
value which is clearly unphysical. A Landau ghost (or pole) is
the name associated with such a singularity. If
\begin{math}\alpha(Q^2)\end{math} can turn negative either by passing
through a zero or through a pole, it would lead to instability
separating a phase where two like charges repel to a different
phase where they attract. In the present work, we shall explore a
general origin of the ghost (and similar pathologies) via the
Dyson equation. While for QED, the phase transition is at
unattainable high energies, in certain condensed matter situations
the singularity is at very low energies and leads to an observable
phase transition called super-radiance. A very practical
application is to the dielectric properties of systems for which a
one loop renormalization leads to Clausius-Mossotti(CM) and
Lorenz-Lorentz(LL) equations. For such phenomenological models,
the phase transition computations can be done to completion
\cite{Siva_2001_1,Siva_2001_2,Siva_con1,Siva_con2}
and we find quantitatively that water is indeed super-radiant at
room temperature, a result predicted earlier by other authors from
somewhat different theoretical considerations\cite{Del_88}.

    In contrast to QED with a Landau ghost, in the same 1-loop
approximation, a non-abelian gauge theory such as QCD for quarks
and glue avoids the Landau ghost (provided the number of quark
flavors is limited to 16) and is asymptotically free (AF).
However, this welcome change in sign of the beta function leading
to AF brings with it its own pathology. The Landau pole is now no
longer a pole on the real axis in the physical region
(\begin{math}s = -Q^2 > 0\end{math}) since it acquires an
imaginary part, but the  sign of the imaginary part is {\it wrong}
i.e., it is negative. It seems appropriate to call this  latter
phenomenon an ``anti-ghost'' since, as we shall see later it also
signals an instability. Physically, it leads to the pleasing
result\cite{Srivastava_2001} that quark and gluon states can only
be transient realized under restricted conditions (of large
energies and momentum transfers). We shall not consider here how
color confinement with normal hadronic phase prevails at low
energies and momentum transfers.

    The paper is organized as follows. In Sec.2, the genesis of
the singularity is considered through a generic discussion of
``feedback'' which is at the heart of the Dyson equation.
In Sec.3, we first develop the CM and LL equations for dielectrics in
order  to exhibit the resulting phase transition as critical
values (of the volume and/or the temperature) leading to
super-radiance are reached for the important case of water (and
other polar materials). In Sec.4, we briefly discuss QCD and
what the anti ghost entails for the resulting physics there.
Sec.5 closes the paper with some concluding remarks and directions for
future work.

\section{Dyson equation and the critical feedback}

\begin{figure}[htbp]
\begin{center}
\mbox{\epsfig{file=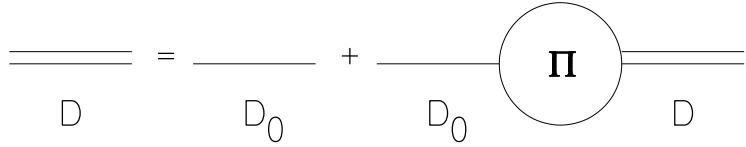,width=10cm}}
\caption{The Dyson equation for the exact propagator $D$
is shown as a linear feedback to the free propagator $D_0$
through the ``self energy'' term $\Pi$.}
\label{lfig1}
\end{center}
\end{figure}

Let \begin{math} D\end{math} be the exact photon propagator,
\begin{math} D_0 \end{math} the free photon propagator and
\begin{math} \Pi \end{math} its proper self energy part\cite{Bjorken}
(suppressing all Lorentz indices and momentum labels momentarily).
The Dyson equation\cite{Dyson_49} then reads
\begin{equation}
 D = D_0 + D_0 \Pi  D = (1 - D_0 \Pi)^{-1}D_0, \label{Dy1}
\end{equation}
or, equivalently written in terms of the inverse propagators
\begin{equation}
 D^{-1} = D_0^{-1} - \Pi . \label{Dy2}
\end{equation}
For computations to follow, it would be useful to recast the logic
behind Eq.(\ref{Dy1}) in the language of linear feedback. Suppose
an external current \begin{math}\delta J_{ext}\end{math} is
applied to a system and we wish to find the corresponding change
in the potential from \begin{math} A \end{math} to \begin{math}A +
\delta A\end{math}. The potential change \begin{math}\delta
A\end{math} is given in terms of the total change in the current
\begin{math}\delta J\end{math} through the free propagator
\begin{math}D_0\end{math} by
\begin{equation}
\delta A = D_0 \delta J = D_0 \left(\delta J_{ext} + \delta
J_{ind}\right) , \label{Dy3}
\end{equation}
where in Eq.(\ref{Dy3}), the total change in the current contains
a feedback current \begin{math}\delta J_{ind}\end{math} in
addition to the applied \begin{math}\delta J_{ext}\end{math}.
The induced piece is the system's linear response to the change
in the potential through the ``self energy'' term
\begin{math}\Pi\end{math}
\begin{equation}
\delta J_{ind} = \Pi \delta A.
\label{Dy4}
\end{equation}
Now defining the exact propagator through
\begin{equation}
\delta A =  D \delta J_{ext}, \label{Dy5}
\end{equation}
Eq.(\ref{Dy1}) follows from Eqs.(\ref{Dy3}-\ref{Dy5}). Eq.(\ref{Dy1}),
when depicted pictorially as in Fig.[\ref{lfig1}],
makes abundantly clear the linear feedback mechanism
implicit in the Dyson equation.

Restoring explicitly the Lorentz indices, we may write the
Feynman gauge, renormalized photon propagator of 4- momentum
\begin{math}q\end{math}
with
\begin{eqnarray}
s = q^2 = (\omega/c)^2 - |\bf k|^2
\nonumber
\end{eqnarray}
(\begin{math}\omega\end{math} the
photon frequency and \begin{math}{\bf k}\end{math}
its wave-vector), in terms of the ``running coupling constant''
\begin{math}\alpha(s)\end{math} and the dielectric
``tensor'' \begin{math}\epsilon(s)\end{math} as follows:
\begin{equation}
\alpha_0  D^{\mu\nu}(q) = \alpha(s) D_0^{\mu\nu}(q) = \frac
{\alpha_0 D_0^{\mu\nu}(q)}{\epsilon(s)}. \label{Dy6}
\end{equation}
Clearly, both poles and zeroes in \begin{math}\alpha\end{math} (which become
respectively, the zeroes and poles of \begin{math}\epsilon\end{math}) are
pathological.

For a renormalization at an arbitrary space-like point
(\begin{math}s = -\Lambda^2\end{math}),
\begin{equation}
\Pi_{\mu\nu}(q) = [-4\pi\alpha(-\Lambda^2)] (q_\mu q_\nu-
\eta_{\mu\nu}q^2) \pi_c(s).
\label{Dy7}
\end{equation}
With \begin{math} \pi_c(s=-\Lambda^2) = 0 \end{math} we have
\begin{equation}
\alpha(s) = \left[\frac {\alpha(-\Lambda^2)}{1 + 4\pi\alpha(-\Lambda^2)
\pi_c(s)}\right]
\label{Dy8}
\end{equation}
and likewise
\begin{equation}
\epsilon(s) = 1 + 4\pi \alpha(-\Lambda^2)\pi_c(s).
\label{Dy9}
\end{equation}

Now let us consider the allowed singularities in the exact
propagator \begin{math} D_{\mu\nu}(q)\end{math}. There are branch
point singularities in \begin{math} D \end{math} (for time-like
momenta at all production thresholds) through unitarity as
\begin{math}\pi_c(s)\end{math} develops an imaginary part. In the
physical region of time-like momenta, there may even be poles
corresponding to resonances on the unphysical Riemann sheet(s). On
the other hand, for space-like momenta,
\begin{math} D \end{math} is forbidden to have any poles,
i.e., on general grounds, we must require that
\begin{equation}
4\pi\alpha(-\Lambda^2)\pi_c(s) > -1, \ \ {\rm  all}\ \ s<0,
\label{Dy10}
\end{equation}
(all space-like momenta for which
\begin{math} \omega < c |{\bf k}| \end{math}).
Whenever this inequality is not satisfied, we shall have an
unphysical Landau pole. The feedback is then beyond its critical
value invalidating linear response and it would signal a phase
transition to another, super-radiant phase of the theory\cite{Siva_2001_2}.

For example, in  one fermion loop QED, renormalized on shell at
\begin{math}s = 0\end{math}, the standard vacuum result reads
\begin{equation}
\epsilon_{1-loop}(s) = 1 - \frac{2\alpha}{\pi}\int_0^\infty
\left(z-z^2\right)
\ln \left[1 - \frac{sz(1-z)}{\kappa^2 - i0^+}\right] dz,
\label{Dy11}
\end{equation}
where \begin{math}\kappa = mc/\hbar \end{math},
with \begin{math}m\end{math} the fermion mass. In the
space-like region \begin{math}\epsilon\end{math} decreases
to zero at the Landau value
\begin{equation}
-s_L = Q_L^2 = \kappa^2 e^{3\pi/\alpha},
\label{Dy12}
\end{equation}
beyond which it turns negative. Thus, QED is in its normal
(stable) phase until an enormously large \begin{math} Q^2
\end{math} value above which it would turn super-radiant. In the
condensed matter example treated in the next section, a similar
phenomenon would occur (this time through a zero in \begin{math}
\alpha \end{math}) but at physically relevant energies and hence
of practical importance.

A discussion of the complementary notion of the anti-ghost is
better done via a subtracted dispersion relation for \begin{math}\epsilon
\end{math}
\begin{equation}
\epsilon(s) = 1 + \left(\frac{s + \Lambda^2}{\pi}\right)\int_0^\infty
\frac {(ds') \Im m\ \epsilon(s')}{(s' + \Lambda^2)(s' - s - i0^+)} ,
\label{Dy13}
\end{equation}
In spinor QED, where only fermion loops contribute,
\begin{math}\Im  m\ \epsilon(s)\end{math} is positive definite
( for all physical \begin{math}s\end{math}). In one
loop, it is given by
\begin{equation}
\Im m\ \epsilon(s) =\frac{ \vartheta(s - 4\kappa^2)
\alpha}{3} \left[1 +
\frac{2\kappa^2}{s}\right]\sqrt{\left[1 - \frac{4\kappa^2}{s}\right]},
\label{Dy14}
\end{equation}
thus the system is ``dissipative''. Whenever the system is
dissipative, the subtraction tells us ( in 1-loop) that there is a
Landau ghost in the space like region.

On the other hand, for QCD, along with the fermion loops (quarks
in this case) as above, which are dissipative, we also have a
larger imaginary part with a negative sign from the gluons
\begin{equation}
\Im m\ \epsilon(s) =\Im m\ \epsilon_{glue}(s) +\Im m\  \epsilon_{quark}(s),
\label{Dy15}
\end{equation}
where
\begin{equation}
\Im m\ \epsilon_{glue}(s) = - \vartheta(s)
\left(\frac{11\alpha_s}{8}\right),
\label{Dy16}
\end{equation}
and
\begin{equation}
\Im m\ \epsilon_{quark}(s)= \sum_{i=1}^{N_f}\vartheta(s -
4\kappa_i^2) \left(\frac{\alpha_s}{12}\right)
\left[1 + \frac{2\kappa_i^2}{s}\right]
\sqrt{\left[1 - \frac{4\kappa_i^2}{s}\right]}
\label{Dy17}
\end{equation}
For the number of flavors \begin{math}N_f < 16\end{math},
the total imaginary part remains negative definite.
This cures the Landau ghost problem and
as well known we have asymptotic freedom, for in the space-like
region \begin{math}s =-Q^2 < - \Lambda^2\end{math},
\begin{equation}
\epsilon(-Q^2)= 1 + \frac{(Q^2 - \Lambda^2)}{\pi} \int_0^\infty
\frac{ds\left[- \Im m\ \epsilon(s)\right]}{(s + \Lambda^2)(s + Q^2)}>1.
\label{Dy18}
\end{equation}
The price paid for exorcizing the Landau ghost into a
benevolent anti-ghost is to turn the system into being
non-dissipative i.e., into an ``amplifier''\cite{Srivastava_2001}
which is well known to be transient.
We shall return to this problem in Sec.4.

\section{Landau ghost in condensed matter and super-radiance}

As stated earlier, unlike the Landau ghost in QED, in condensed
matter we have unphysical singularities at low frequencies which
signal phase transitions of practical importance for the physics
and the chemistry of substances.

In condensed matter literature, the analogue of
\begin{math}\alpha \pi_c\end{math}
in Eq.(\ref{Dy8}) is known as the electric susceptibility
\begin{math}\chi_e\end{math} and
the object here is to relate it to macroscopic quantities such as
the polarizability of the molecules \begin{math}\hat \alpha\end{math}
and \begin{math}v \end{math} the
volume per molecule.( We have defined the polarizability with a
``hat'' in order to distinguish it from the coupling constant
\begin{math}\alpha\end{math}).

The CM equation\cite{Mossotti_50,Clausius_97} is an ideal example
of a linear feedback paralleling the argument of the last section.
To briefly recall what is involved and to set the notation, the
electric dipole moment per unit volume or, polarization
\begin{math}{\bf P}\end{math}
\begin{equation}
{\bf P} = \chi_e {\bf E} = N \left<{\bf p}_{mol}\right>,
\label{Lg1}
\end{equation}
where \begin{math} N \end{math} is the number of molecules per
unit volume and \begin{math}\left<{\bf p}_{mol}\right>\end{math}
is the mean dipole moment of the molecules. Then, without the
``feedback'', justifiable in rare media where molecular separations
are large, one assumes that the macroscopic field
\begin{math}{\bf E}\end{math} is the same as that acting on
a given molecule
\begin{equation}
\left<{\bf p}_{mol}\right> = {\hat \alpha} {\bf E},
\label{Lg2}
\end{equation}
in terms of the polarizability. So
that for rare  media
\begin{equation}
\chi_e = N {\hat \alpha} = {\hat \alpha}/v,
\label{Lg3}
\end{equation}
where \begin{math} v \end{math} is the volume occupied by a single molecule.

In denser media where molecular separations are small,
the polarization \begin{math}{\bf P}\end{math} induces an
internal electric field \begin{math}{\bf E}_i\end{math}. If one includes
only the contributions from molecules far away, this feedback
term is estimated to be
\begin{equation}
{\bf E}_i = \left(\frac{4\pi}{3}\right) {\bf P}.
\label{Lg4}
\end{equation}
Now, Eq.(\ref{Lg1}) leads to
\begin{eqnarray}
{\bf P} &=& N \left<{\bf p}\right> = N {\hat \alpha}
\left({\bf E} + {\bf E}_i\right)\nonumber \\
& =& N {\hat \alpha} \left({\bf E} +
\frac{4\pi}{3} {\bf P}\right)\nonumber \\
&=& \left[\frac{N\hat\alpha}{1 -
(4\pi /3)N{\hat \alpha}}\right]{\bf E}.
\label{Lg5}
\end{eqnarray}

It is perhaps not surprising therefore, that the CM
equation gives the same expression for
\begin{math}\chi_e\end{math} as is obtained
from the lowest order mean field theory\cite{Schan_92,Fu_93}
\begin{equation}
\chi_e = \left(\frac{({\hat \alpha}/v)}{1 - (4\pi{\hat
\alpha}/3v)}\right),
\label{Lg6}
\end{equation}
and for \begin{math}\epsilon \end{math} one finds

\begin{equation}
\epsilon = \left(\frac{1 + 2(4\pi{\hat \alpha}/3v)}{1 -
(4\pi{\hat \alpha}/3v)}\right).
\label{Lg7}
\end{equation}
The normal phase would require for its stability that the static
polarizability satisfy
\begin{equation}
4\pi {\hat \alpha} < 3v,
\label{Lg8}
\end{equation}
in order for no poles to develop in
\begin{math}\epsilon \end{math} (or \begin{math}\chi_e \end{math}). On
the other hand, as the molecular density decreases below a
critical value \begin{math}v_c \end{math}, i.e., for
\begin{equation}
v  < v_c = \left(\frac{4\pi}{3}\right) {\hat \alpha},
\label{Lg9}
\end{equation}
the normal phase would become unstable yielding to a new stable
super-radiant phase of Dicke-Lieb-
Preparata\cite{Hepp_73,Wang_73,Emelanov_76,Prep_90,Del_98} models.

CM equations become the Lorenz-Lorentz (LL) equations
\cite{Lorenz_80,Lorentz_80,Souk_94,Ersfeld_98}
as static \begin{math}{\hat \alpha},\  \chi_e\ {\rm and}\
\epsilon \end{math} get generalized
to a finite frequency \begin{math}\omega \end{math}. Here
one finds giant (macroscopic)
Lamb shifts in frequency due to the ``feedback''. Let us suppose
that a single atom with atomic number \begin{math}Z\end{math} has
a strong resonance
at a frequency \begin{math}\omega_\infty\end{math}, so that
\begin{equation}
{\hat \alpha}(\omega) \approx
\left(\frac{Ze^2}{m}\right)\frac{1}
{\omega_\infty^2 - \omega^2 - 2i\gamma\omega},
\label{Lg10}
\end{equation}
where \begin{math} m \end{math} is the electron mass.
Eqs(\ref{Lg6}) and (\ref{Lg10}) then imply
\begin{equation}
\chi_e(\omega) \approx
\left(\frac{Ze^2}{m}\right)\frac{1}{\omega_0^2 -
\omega^2 - 2i\gamma\omega},
\label{Lg11}
\end{equation}
where the ``renormalized'' resonant frequency
\begin{math}\omega_0\end{math} is given
by
\begin{equation}
\omega_0^2 = \omega_\infty^2 - \left(\frac{4\pi Ze^2}{3mv}\right) =
\omega_\infty^2\left[ 1 - (\frac{4\pi {\hat \alpha}}{3v})
\right],
\label{Lg12}
\end{equation}
where \begin{math}{\hat \alpha} \equiv
{\hat \alpha}(\omega = 0)\end{math}

At the critical temperature \begin{math}T_c\end{math},
the static polarizability stability condition given by
Eq.(\ref{Lg8}) becomes
\begin{equation}
\left(\frac{4\pi {\hat \alpha}(T_c)}{3}\right)
= v\ \ \ ({\rm critical\ point}).
\label{Lg13}
\end{equation}
Dynamically, the renormalized frequency in Eq.(\ref{Lg12}) is driven to
zero at the critical temperature
\begin{equation}
\omega_0\left(T\to T_c+0^+\right)=0,
\label{Lg14}
\end{equation}
which again leads to Eq.(\ref{Lg13}).

In previous work\cite{Siva_con1}, we have obtained the phase diagram
through (2 and 4 level) model Hamiltonians  which allow us to
compute
\begin{math} {\hat\alpha}(T)\end{math} for a single molecule. It
is to be emphasized that perturbative corrections would shift the
exact boundary of the super-radiant phase from its CM value but
its occurrence is left intact.

Here we simply estimate the polarizability of water
using the Debye model\cite{Debye}
\begin{equation}
{\hat \alpha} = {\hat \alpha}_o + \left(\frac{\mu^2}{3k_B
T}\right).
\label{Lg15}
\end{equation}
For water\cite{Water1,Water2}
\begin{equation}
{\hat \alpha}_0\ =\ 1.494 A^3 \ \ {\rm and} \ \ \mu = 1.855\times
10^{-18} Gauss\ cm^3.
\label{Lg16}
\end{equation}
This gives for the polarizability of water at room temperature
\begin{equation}
{\hat \alpha}_{water} \approx 28.6962 \AA^3.
\label{Lg17}
\end{equation}
From the molecular density of water, we find
\begin{math}v \approx\ 30.0137 \AA^3\end{math},
so that the super-radiance condition
\begin{equation}
\frac{v_c}{v} = \left(\frac{4\pi {\hat \alpha}}{3v}\right)
\approx 4.0054 > 1,
\label{Lg18}
\end{equation}
is satisfied showing that water is indeed super-radiant at room
temperature. Surely we expect numerical corrections to the above
result. But the ``margin of safety'' is reasonable so that the
claim about room temperature water being super-radiant is robust.

\begin{figure}[htbp]
\begin{center}
\mbox{\epsfig{file=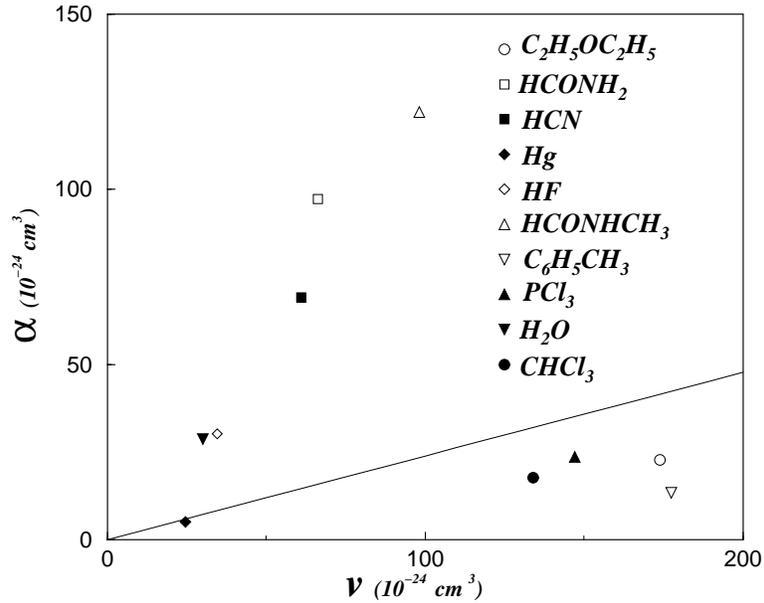,width=10cm}}
\caption{Polarizabilties ${\hat \alpha}$ for some materials are shown as a
function of the volume per molecule $v$. Super-radiant materials lie
above the line shown with slope [$3/(4\pi)$].}
\label{lfig2}
\end{center}
\end{figure}

To show the efficacy of the above criterion, we have carried a
similar analysis for a large number of polar and non-polar
materials\cite{CRC}. The results shown in Fig.[\ref{lfig2}]
confirm that at room temperatures, many polar liquids are
super-radiant whereas non-polar materials are not.

\section{Implications of the anti-ghost in QCD}

In this section, we resume our discussion of the physical
implications arising from the negative sign of
\begin{math}\Im m\ \epsilon\end{math}
computed in 1-loop QCD. In general, such a negative sign signals
a lack of causality, the latter driving the system to become an
``amplifier'' instead of a ``dissipator''. An amplifier (for
example a laser with an inverted population entails the notion
of a negative temperature) can only be a transient phenomenon since
it requires a source to feed energy into it.

We may conclude from the above that the \begin{math}J = 1\end{math},
color octet channel of QCD represented by a gluon propagator with a
negative absorptive part cannot be a part of the physical spectrum.
We may conjecture that absorptive parts for all channels carrying
color are of the amplifier variety thereby excluding all such states
from the physical spectrum. At best then quarks and gluons can occur
only transiently as virtual states in a restricted part of the phase
space spanned by hadrons, the colorless children of quarks
and glue. Only after one is able to  produce hadrons as colorless
composites of quarks and gluons, would it be possible to check
whether these true inhabitants of the ``normal'' phase of QCD
have dissipative absorptive parts. Needless to say that we are
nowhere near that goal.
\section{Conclusions and outlook}
We have presented a general discussion of the Landau ghost
and related phenomena through the Dyson equation and these pathologies
have been found to trigger phase transitions. A convincing case has been
made for polar liquids to be in the super-radiant phase at ambient
temperatures substantiating previous work on water. Similar arguments
for QCD, through a wrong sign absorptive part in the gluon propagator,
show that the gluons (and conjectured more generally for any channel
carrying color) cannot be a part of the physical spectrum. This
generalization needs verification. Also, further work is needed to
show that the spectral functions for hadrons (that is the color
singlet bound states) in QCD do possess positive absorptive parts
as they must to qualify as the true ``normal'' ground states of the
theory.

\end{document}